\documentclass[twocolumn]{aastex701}

\usepackage{amsmath}	
\usepackage{amssymb}	
\usepackage{booktabs}
\usepackage{lipsum}
\usepackage[T1]{fontenc}
\usepackage{fix-cm}
\usepackage{graphicx}	

\usepackage{newtxtext,newtxmath}
\usepackage{natbib}

\let\tablenum\relax
\usepackage{siunitx}
\usepackage{tikz}
\usepackage{float}
\usepackage{CJK}

\newcommand{\Gaia}{\textit{Gaia}}
\newcommand{\TESS}{\textit{TESS}}
\newcommand{\kapvel}{$\kappa$ Vel}
\newcommand{\chinese}[1]{\begin{CJK}{UTF8}{gbsn}#1\end{CJK}}

\begin{document}

\title{Hidden in Plain Sight II: Characterizing the luminous companion to $\kappa$ Velorum with VLTI/GRAVITY}

\author[0000-0003-2431-981X]{D. M. Rowan}\thanks{NHFP Hubble Fellow}
\email{dmrowan@berkeley.edu}
\affiliation{Department of Astronomy, University of California, Berkeley, CA 94720, USA}

\author{S. Kraus}
\affiliation{Astrophysics Group, Department of Physics \& Astronomy, University of Exeter, Stocker Road, Exeter EX4 4QL, UK}
\email{S.Kraus@exeter.ac.uk}

\author[0000-0003-2377-9574]{Todd A. Thompson}
\affiliation{Department of Astronomy, The Ohio State University, 140 West 18th Avenue, Columbus, OH, 43210, USA}
\affiliation{Center for Cosmology and Astroparticle Physics, The Ohio State University, 191 W. Woodruff Avenue, Columbus, OH, 43210, USA}
\affiliation{Department of Physics, The Ohio State University, Columbus, Ohio, 43210, USA}
\email{thompson.1847@osu.edu}

\correspondingauthor{D. M. Rowan (dmrowan@berkeley.edu)}



\begin{abstract}

\kapvel{} (Markeb, HD 81188, \chinese{天社五}) is one of the brightest stars in the Southern sky and has long been known to be a single-lined spectroscopic binary. The binary mass function is large, $f(M)=1.15\ M_\odot$, suggesting that the bright ($V=2.5$) B2IV star may host a dark, compact object companion. We use VLTI/GRAVITY observations to definitively test this possibility by directly resolving the binary. We detect a main sequence B star companion and rule out the compact object scenario. By combining the relative astrometric orbit and archival radial velocities, we report an updated precise characterization of the orbit (period $P=116.795\pm0.002$~d, eccentricity $e=0.1764\pm0.0004$, inclination $i=74.04\pm0.01^{\circ}$) and estimate the masses of the B stars. Using the original Hipparcos parallax measurement $\varpi = 6.05\pm0.48$~mas, we find $M_1 = 10^{+4}_{-2}\ M_\odot$ and $M_2 = 6.9\pm1.0\ M_\odot$. The uncertainties on the masses are primarily driven by the uncertain parallax, which we find is likely biased by the orbital motion. We use an archival UVES spectrum and MIST evolutionary tracks to refine our mass estimates. Finally, we discuss how interferometry and high-contrast imaging may be used to characterize other candidate star+compact object binaries, including those that will be discovered with \Gaia{} DR4, as part of a larger effort to uncover the hidden population of black holes in the Milky Way. 

\end{abstract}



\section{Introduction} \label{sec:intro}

The majority of the observed black holes and neutron stars in the Milky Way have been detected in X-ray binary systems \citep{CorralSantana16}. While compact object masses can sometimes be measured dynamically in these systems, either by inferring the orbital inclination from the accretion disk properties \citep[e.g.,][]{Torres20} or by modeling ellipsoidal modulations \citep[e.g.,][]{Orosz01}, many of the transient BH candidates have no mass measurements \citep[e.g.,][]{Russell22}. Since X-ray binaries are expected to undergo at least one episode of binary interaction and mass transfer \citep{Kalogera96, Tauris06}, the observed compact object mass distribution from X-ray binaries may present a biased view of the intrinsic distribution.

Outside of the Galaxy, gravitational waves have been used to detect $>150$ compact object binary mergers \citep{Ligo2025}. While these targets provide powerful constraints on the mass distribution of black holes \citep{Ligo2025_population}, these rare extragalactic systems are also the products of complex binary interactions that result in a coalescence. The majority of the observed gravitational wave sources are also more massive than typical X-ray binaries, since more massive black hole mergers have a larger characteristic strain. 

In recent years, there has been an increased effort to uncover the hidden population of dormant black holes in the Milky Way. Isolated black holes can only be detected via microlensing \citep{Lam22, Lam23}. While there is currently only one strong candidate, the Vera C. Rubin Legacy Survey of Space and Time and the Roman Space Telescope are expected to detect $\gtrsim 30$ isolated black holes \citep{Abrams25, Lam23_whitepaper}. Multiple observational methods can be used to detect and characterize black holes in wide binaries with luminous companions. Three black holes have been detected using \Gaia{} astrometry \citep{ElBadry23_BH1, ElBadry23_BH2, Chakrabarti23, Tanikawa23}, including \Gaia{}~BH3, a red giant star in the Milky Way halo with a $33\ M_\odot$ black hole in an 11.9~year binary orbit \citep{GaiaBH3}. Spectroscopic surveys have also been applied to search for candidates \citep[e.g.,][]{Giesers18, Shenar22, Mahy22}, but the unknown orbital inclination means that only a minimum black hole mass can be measured. At short orbital periods ($P \lesssim 10$~days for main sequence companions), ellipsoidal modulations can be used to measure the orbital inclination \citep{Morris93}, but no strong candidates have been discovered starting from photometric variability \citep{Rowan21, Green25}.

Regardless of whether a black hole candidate is detected with astrometry, spectroscopy, or photometry, confirming a black hole effectively means ruling out any type of luminous companion. For some systems, like the $33\ M_\odot$ black hole in \Gaia{}~BH3, the spectral energy distribution can easily exclude a massive main sequence star or an equal-mass binary of two $16.5\ M_\odot$ stars. For lower-mass black holes, spectral disentangling can help to distinguish compact objects from false positives, such as rapidly rotating subgiants \citep[e.g.,][]{ElBadry22_zoo, Jayasinghe22}. However, if we can directly resolve a candidate system with high-contrast imaging or interferometry, we can detect or rule out luminous companions with a single observation \citep[e.g.,][]{Frost22}. 

If we observe a stellar binary, the direct imaging observations will resolve the two stars and help constrain the three-dimensional orbit. If we instead resolve a non-interacting black hole binary, we expect to observe a single point source. The combination of RVs and astrometry can therefore be used to unambiguously detect and characterize non-interacting black hole binary systems. In \citet{Rowan25_shark}, we performed a pilot study to explore the feasibility of this method using SHARK-VIS \citep{Pedichini22, Pedichini24} on the Large Binocular Telescope \citep{Hill10}. We ruled out black holes for three of the four systems observed, and showed that two of the binaries selected are likely hierarchical triples. Since the SHARK-VIS observations only require $\sim 10$~min of on-sky observing time, direct imaging can be used to rapidly vet black hole candidates in long period orbits more efficiently than high-resolution spectroscopic follow-up. 

Here, we use VLTI/GRAVITY to observe $\kappa$ Velorum (hereafter \kapvel{}), a bright ($V=2.5$), single-lined spectroscopic binary with an RV orbit that indicates a large, massive companion. Section \S\ref{sec:targetselection} describes how we selected this target as a potential black hole candidate. Section \S\ref{sec:gravity} reports the VLTI/GRAVITY observations and the describes the data reduction. We unambiguously detect a second star in the GRAVITY observations and rule out a black hole. We then characterize the stellar binary by simultaneously fitting the archival RVs with the relative astrometry in Section \S\ref{sec:orbit}, model an archival high resolution spectrum (Section \S\ref{sec:uves}), and make comparisons to evolutionary tracks of massive stars (Section \S\ref{sec:evotracks}). Finally, we discuss the prospects for observing other high-mass function single-lined spectroscopic binaries with interferometry and direct imaging in Section \S\ref{sec:discussion}.

\section{Target Selection} \label{sec:targetselection}

\begin{figure}
    \centering
    \includegraphics[width=\linewidth]{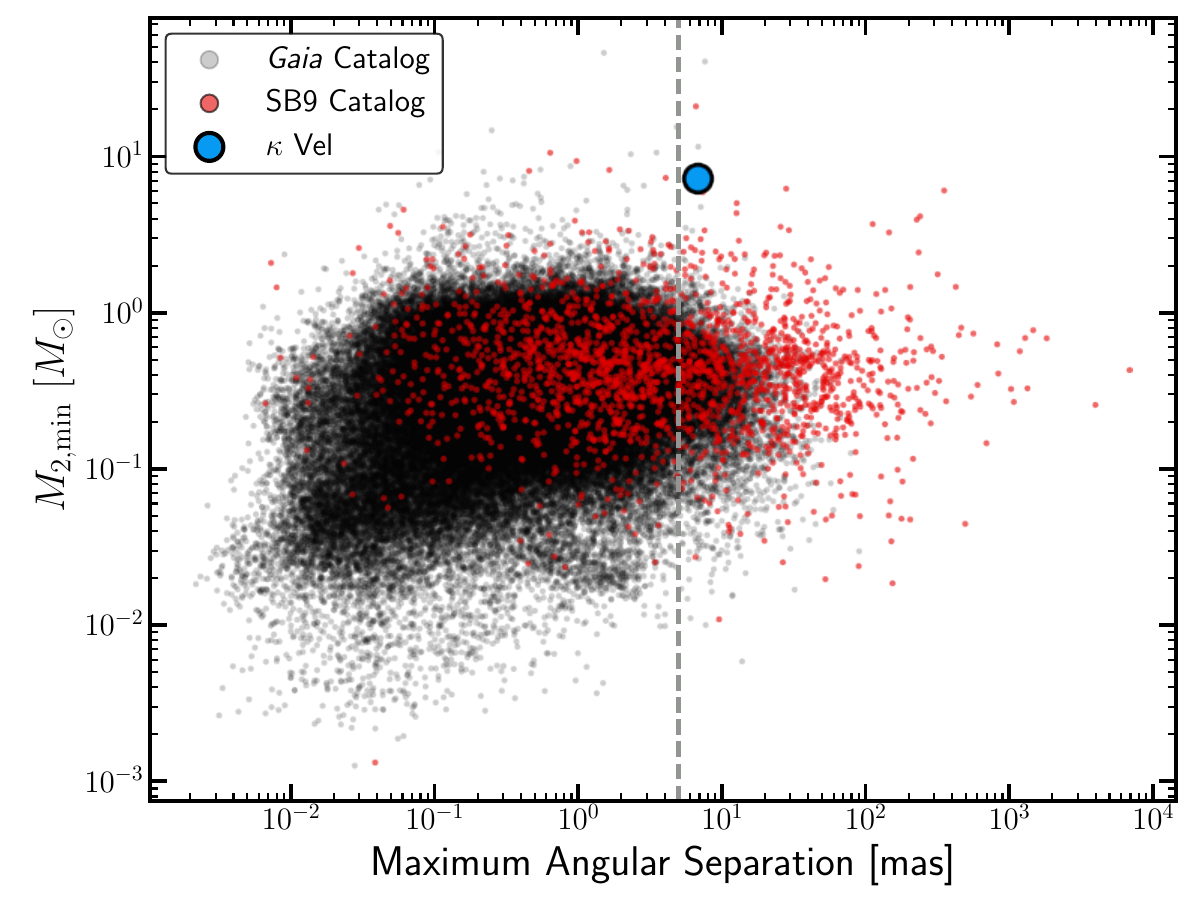}
    \caption{Minimum companion masses, $M_{2,\rm{min}}$, and maximum projected angular separation for binaries in the \Gaia{} SB1s and SB9 catalogs. \kapvel{} is marked in blue. The vertical line marks the projected angular separation resolvable with VLTI/GRAVITY.}
    \label{fig:candidate_selection}
\end{figure}

We selected targets following the procedure outlined in \citet{Rowan25_shark}. In short, we started from two large catalogs of spectroscopic binaries, the Ninth Catalog of Spectroscopic Binary Orbits \citep[][hereafter SB9]{Pourbaix04} and single-lined spectroscopic binaries from the \Gaia{} DR3 non-single stars catalog \citep[][hereafter \Gaia{} SB1]{GaiaHiddenTreasure, Gosset25}. We used the astrophysical parameter estimates from {\tt StarHorse} \citep{Anders22} to estimate the mass of the photometric primaries, and determined the minimum companion masses and maximum projected orbital separations based on the archival RV orbits. Figure \ref{fig:candidate_selection} shows the maximum projected orbital separations and minimum companion mass measurements.

\begin{figure}
    \centering
    \includegraphics[width=\linewidth]{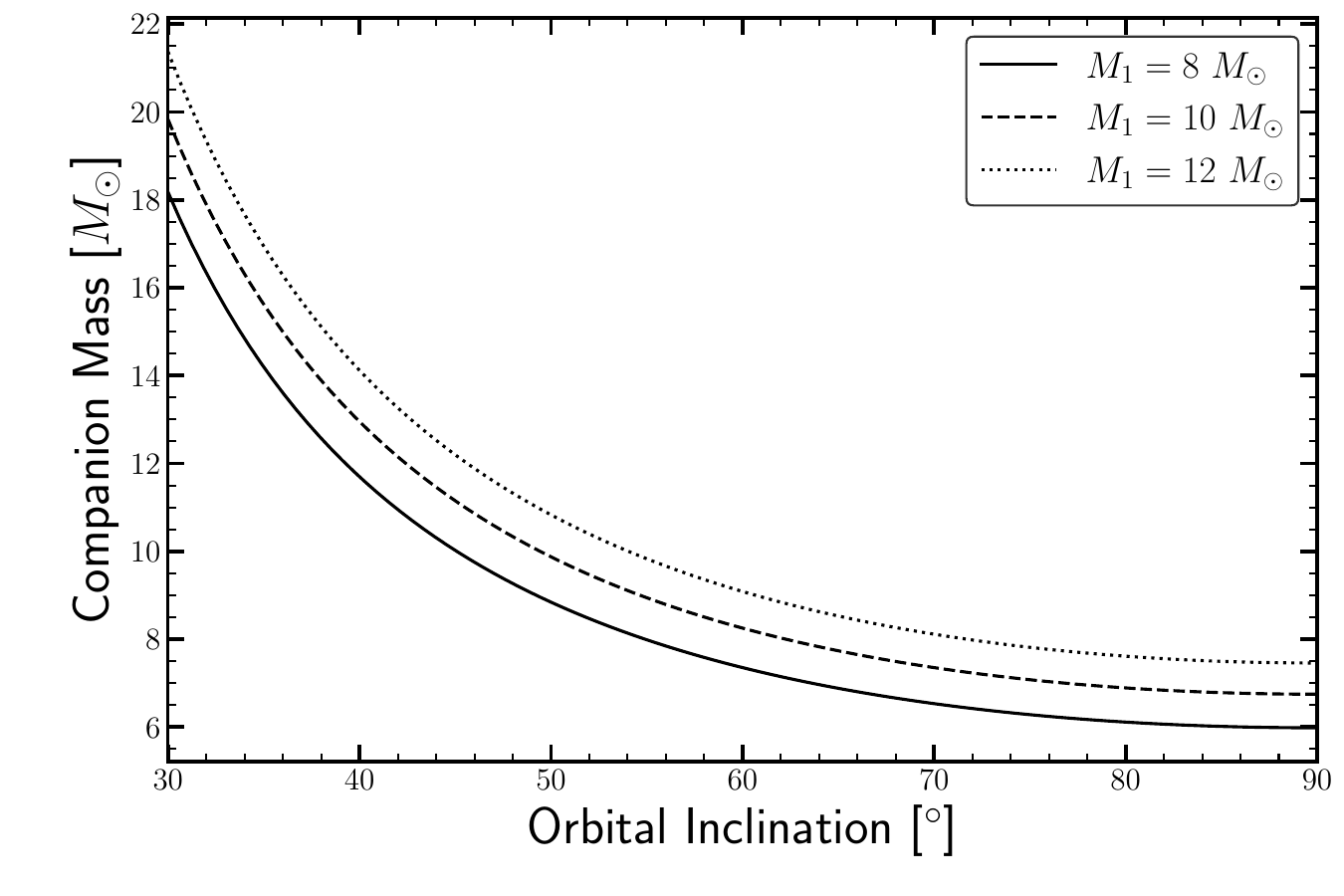}
    \caption{Companion mass, $M_2$, as a function of the orbital inclination for different values of the primary mass, $M_1$. Even if the primary mass is $8\ M_\odot$, the companion must be $\gtrsim 6\ M_\odot$.}
    \label{fig:companion_mass}
\end{figure}

We identified \kapvel{} (HD 81188, HIP 45941) as having a large minimum companion mass and an angular separation accessible with VLTI/GRAVITY. With an apparent magnitude of $V=2.5$, \kapvel{} is one of the brightest stars in the Southern sky and has been targeted as part of a number of surveys, but the secondary has not previously been characterized. \kapvel{} has been most commonly observed as part of interstellar medium absorption line studies \citep[e.g.,][]{Crawford91, Fruscione94, Jenniskens97, Crawford02, Smith13, Cox17}. \kapvel{} was first identified as a binary candidate by \citet{Wright05} and the orbit was characterized by \citet{Curtis1907} as part of the D.O. Mills Expedition from Lick Observatory to the Southern Hemisphere. Five additional measurements taken by \citet{Buscombe60} confirmed the RV variability and orbit. Table \ref{tab:archival_rvs} lists the archival measurements. We assume an RV uncertainty of 5~km/s for the \citet{Curtis1907} measurements since measurement errors were not reported. \citet{Chini12} also measured four RVs of \kapvel{} as part of a spectroscopic surveys of massive stars. They classify the system as an SB1, but the individual RV measurements or spectra are not available. \kapvel{} is included in the SB9 catalog with an orbital period of $P=116.65$~days, $K_1=46.5$~km/s, and a binary mass function $f(M)=1.15\ M_\odot$, where
\begin{equation} \label{eqn:massfunction}
    f(M) = \frac{P K^3}{2\pi G}\left(1-e^2\right)^{3/2} = \frac{M_2^3 \sin^3i}{(M_1+M_2)^2},
\end{equation}
$M_1$ is the photometric primary mass, $M_2$ is the companion mass, $i$ is the orbital inclination, and $e$ is the orbital eccentricity. The photometric primary is spectroscopically characterized as a B2IV star \citep{Levato75}, which roughly corresponds to a primary mass of $\sim 7$--$10\ M_\odot$ \citep{Pecaut13}. Based on the RV orbit, even if the primary is $7\  M_\odot$ and the orbit is edge-on, the companion must be $>5.7\ M_\odot$ (Figure \ref{fig:companion_mass}). Because of the large minimum companion mass,  \citet{Trimble69} included \kapvel{} in their catalog of binaries that could host compact object companions. The companion could be another massive star, a compact object, or an inner binary of two lower mass stars. The maximum projected orbital separation is $>7.0$~mas (Figure \ref{fig:candidate_selection}), making this a suitable target for VLTI direct imaging observations to test these possibilities. 

\begin{table}
    \centering
    \caption{Archival RV measurements of \kapvel{} from \citet{Curtis1907} and \citet{Buscombe60}. Since \citet{Curtis1907} do not report RV uncertainties, we estimate a conservative RV uncertainty of 5 km/s. We also include our RV measurement from the UVES spectrum. We assign an RV uncertainty on the UVES spectrum of $2$~km/s to account for the unknown systematic RV offset between the archival measurements and the UVES measurement.}
    \begin{tabular}{l r r r}
\toprule
{$\rm{JD}-2.4165\times10^6$} & {RV} & {$\sigma_{\rm{RV}}$} & {Reference} \\
{[d]} & {[km/s]} & {[km/s]} & {}\\ 
\midrule
46.24 & 68.4 & 5.0 & Curtis07 \\
360.20 & 12.9 & 5.0 & Curtis07 \\
397.15 & 65.7 & 5.0 & Curtis07 \\
412.10 & 53.3 & 5.0 & Curtis07 \\
1087.35 & 58.6 & 5.0 & Curtis07 \\
1088.29 & 57.9 & 5.0 & Curtis07 \\
1090.33 & 58.5 & 5.0 & Curtis07 \\
1091.33 & 64.8 & 5.0 & Curtis07 \\
1097.29 & 65.8 & 5.0 & Curtis07 \\
1109.29 & 62.0 & 5.0 & Curtis07 \\
1154.04 & -21.0 & 5.0 & Curtis07 \\
1155.06 & -19.2 & 5.0 & Curtis07 \\
1158.07 & -15.2 & 5.0 & Curtis07 \\
1159.05 & -14.5 & 5.0 & Curtis07 \\
1186.09 & 33.8 & 5.0 & Curtis07 \\
1191.07 & 38.2 & 5.0 & Curtis07 \\
1192.06 & 43.2 & 5.0 & Curtis07 \\
1195.98 & 46.7 & 5.0 & Curtis07 \\
1201.00 & 52.7 & 5.0 & Curtis07 \\
1240.97 & 22.1 & 5.0 & Curtis07 \\
1245.96 & 0.3 & 5.0 & Curtis07 \\
1248.97 & -7.6 & 5.0 & Curtis07 \\
1249.98 & -8.8 & 5.0 & Curtis07 \\
1250.96 & -13.3 & 5.0 & Curtis07 \\
1252.96 & -19.2 & 5.0 & Curtis07 \\
1257.95 & -29.0 & 5.0 & Curtis07 \\
1258.96 & -24.5 & 5.0 & Curtis07 \\
18321.98 & -21.0 & 4.5 & Buscombe60 \\
19736.10 & 12.0 & 3.2 & Buscombe60 \\
19760.96 & 39.0 & 7.1 & Buscombe60 \\
20472.10 & 45.0 & 4.9 & Buscombe60 \\
20501.99 & 38.0 & 6.2 & Buscombe60 \\
41361.55 & 26.6 & 2.0 & UVES \\
\bottomrule
\end{tabular}

    \label{tab:archival_rvs}
\end{table}

\section{VLTI/Gravity Observations} \label{sec:gravity}

We observed \kapvel{} at nine epochs between December 2024 and February 2025 using the European Southern Observatory's Very Large Telescope Interferometer (VLTI) as part of observing program 114.274C (PI: Thompson). The GRAVITY instrument \citep{Gravity17} was used to combine the four auxiliary telescopes (ATs) that were in the large (A0-G1-J2-K0) and extended (A0-B5-J2-J6) array configuration, resulting in maximum baselines $D_{\mathrm{max}}$ up to 129.3 and 201.5\,meters, respectively. The observations were conducted in the $K$-band (2.0-2.45\,$\mu$m), resulting in an effective angular resolution of $\lambda/2D_{\mathrm{max}}=1.0$~milliarcseconds (mas). The GRAVITY science combiner spectrograph was configured for medium-spectral resolution (MR, $R=\lambda / \Delta\lambda=500$) for the first five observations to achieve the highest signal-to-noise and the lowest companion detection limits. From these initial observations, it became clear that the system is a relatively low-contrast binary, so we switched to GRAVITY's high-spectral resolution mode (HR, $R=4000$) for the last three observations.  We observed calibrator stars with known diameters that were observed immediately after the science star and used to calibrate instrumental and atmospheric effects. The observing dates, detector integration times (DITs), and choice of calibrator for each night are listed in Table~\ref{tab:obslog}. We used the public GRAVITY pipeline, version 1.7.0 to derive from each data set six calibrated visibilities and four closure phases. Given that the observation used a single calibrator per night, we applied a minimum squared visibility error of 0.02 and a minimum closure phase error of $0.5^{\circ}$ in order to achieve a realistic representation of measurement uncertainties.

\begin{deluxetable*}{cccccccc}
\tablecolumns{8}
\tablewidth{0pc}
\tablecaption{VLTI/GRAVITY observation log. The dates are given for the start of the night. The calibrator uniform-disk (UD) diameters are estimated with the JMMC SearchCal tool \citep{Chelli16}.} \label{tab:obslog}
\tablehead{
\colhead{Date}  & \colhead{MJD} & \colhead{Telescope}     & \colhead{Spectral}  & \colhead{DIT} & \colhead{Calibrator} & \colhead{Calibrator diameter} \\
\colhead{(UT)}  & \colhead{}         & \colhead{configuration} & \colhead{setup} & \colhead{[s]} & \colhead{} & \colhead{[mas]} }
\startdata
2024-12-25    & 60670.24654      & A0-G1-J2-K0  & MR  &  1     & HD81720 & $0.834\pm0.073$ \\
2024-12-28   & 60672.23178       & A0-G1-J2-K0   & MR   & 1    & HD83465 & $0.766\pm0.077$ \\
2025-01-03   & 60679.26416       & A0-B5-J2-J6   & MR   & 1    & HD80093 & $0.720\pm0.071$ \\  
2025-01-08   & 60684.24311       & A0-B5-J2-J6   & MR   & 1    & HD83465 & $0.766\pm0.077$ \\
2025-02-08   & 60715.17032       & A0-B5-J2-J6   & MR   & 1    & HD80765 & $0.827\pm0.080$ \\
2025-02-18   & 60725.17185       & A0-B5-J2-J6   & HR   & 10   & HD81720 & $0.834\pm0.073$ \\
2025-02-21   & 60728.04764       & A0-B5-J2-J6   & HR   & 10    & HD80093 & $0.720\pm0.071$ \\
2025-02-22   & 60729.07190       & A0-B5-J2-J6   & HR   & 10    & HD80765 & $0.827\pm0.080$ \\
2025-02-28   & 60731.05039       & A0-G1-J2-K0   & HR  & 10  & HD83465 &  $0.766\pm0.077$ \\
\enddata
\end{deluxetable*}

\subsection{Astrometry model-fitting} \label{sec:ast_model_fitting}

The visibility and closure phase data shows strong sinusoidal modulations that are indicative of a close binary system.  In order to quantify the binary star parameters, we fit geometric models with the PMOIRED modeling software \citep{Merand22}. The free parameters are the relative separation vector of the two stars measured from the primary to the secondary (dRA, dDEC) and their $K$-band flux ratio ($\epsilon = f_2/f_1$). While we resolve the two stars, we do not resolve their stellar disks, so we fix the disk diameters to $0.1$~mas and use a uniform-disk model. For the error estimation, we bootstrap over the individual measurements that were taken during each epoch. The derived astrometry is reported in Table~\ref{tab:astrometry}.

\begin{deluxetable}{ccccc}
\tablecolumns{5}
\tablewidth{0pc}
\tablecaption{Astrometry for \kapvel{}. The measurement uncertainties are estimated by bootstrapping over the individual measurements taken during each epoch (Section \S\ref{sec:ast_model_fitting}). $\rho$ is the correlation coefficient between the dRA and dDEC measurements and $f_2/f_1$ is the flux ratio in the $K$-band.} \label{tab:astrometry}
\tablehead{
\colhead{Date}  & \colhead{dRA} & \colhead{dDEC}  & \colhead{$\rho$} & \colhead{$f_2/f_1$}  \\
\colhead{(UT)}  & \colhead{[mas]}         & \colhead{[mas]} & \colhead{} & \colhead{} 
}
\startdata
2024-12-25   & $-0.956\pm0.002$ & $-3.737\pm0.003$ & $-0.33$ & $0.28\pm0.01$  \\
2024-12-28   & $-0.678\pm0.003$ & $-4.259\pm0.003$ & $-0.61$ & $0.28\pm0.01$  \\
2025-01-03   & $+0.331\pm0.002$ & $-5.806\pm0.002$ & $-0.17$ & $0.29\pm0.01$  \\
2025-01-08   & $+1.025\pm0.002$ & $-6.568\pm0.001$ & $-0.01$ & $0.29\pm0.01$  \\
2025-02-08   & $+2.203\pm0.003$ & $-1.529\pm0.003$ & $-0.19$ & $0.29\pm0.01$  \\
2025-02-18   & $+0.631\pm0.005$ & $+2.756\pm0.004$ & $-0.22$ & $0.29\pm0.01$  \\
2025-02-21   & $+0.081\pm0.001$ & $+3.833\pm0.001$ & $-0.32$ & $0.28\pm0.01$  \\
2025-02-22   & $-0.122\pm0.002$ & $+4.176\pm0.001$ & $+0.60$ & $0.28\pm0.01$  \\
2025-02-28   & $+0.510\pm0.001$ & $+4.781\pm0.002$ & $-0.50$ & $0.28\pm0.01$  \\
\enddata
\end{deluxetable}

\section{Orbit Fitting} \label{sec:orbit}

We can constrain the binary orbit of \kapvel{} by combining the archival RVs and the relative astrometry. Given an additional measurement of the parallax, the individual component masses can be determined. 

We simultaneously model the RVs and relative astrometry with Markov Chain Monte Carlo (MCMC) with {\tt emcee} \citep{ForemanMackey13}. We sample over six orbital parameters: the period, $P$, the periastron time, $T_0$, the orbital inclination, $i$, the longitude of the ascending node, $\Omega$, the eccentricity, $e$, and the longitude of periastron, $\omega$. We parameterize the latter two as $\sqrt{e}\cos\omega$ and $\sqrt{e}\sin\omega$ for computational efficiency. The angular size of the astrometric orbit is the ratio of the orbital semi-major axis to the distance, $a_0 = a/d$. We sample over the velocity-semi amplitude of the photometric primary, $K_1$, and the system velocity, $\gamma$. We include an offset $\delta_{\rm{RV}}$ between the \citet{Curtis1907} and \citet{Buscombe60} RVs to account for systematic differences between the instruments. We use a Gaussian prior on $\delta_{\rm{RV}}$ centered on zero with $\sigma = 1.5$~km/s. Since there is only one UVES spectrum, we do not fit for the potential RV offset between the \citet{Curtis1907} RVs and the UVES spectrograph, and instead choose to inflate the measured UVES RV uncertainty (see Section \S\ref{sec:uves}). 

Finally, we include two additional terms to account for potentially underestimated uncertainties. The stellar jitter term, $s$, accounts for stellar variability and underestimated RV uncertainties \citep{PriceWhelan17}. To account for underestimated astrometric uncertainties, we also include a multiplicative term, $s_{\rm{ast}}$, that scales the bootstrapped astrometric uncertainties.

The log-likelihood is the sum of the log-likelihoods for the RV and astrometric measurements, 

\begin{equation}
    \ln \mathcal{L} = \ln \mathcal{L}_{\rm{RV}} + \ln \mathcal{L}_{\rm{ast}}.
\end{equation}
\noindent The log-likelihood for the radial velocity observations is 
\begin{multline}
    \ln \mathcal{L}_{\text{RV}} = -\frac{1}{2} \sum_i \Bigg[
    \frac{(\text{RV}_{\text{pred}}(t_i) - \text{RV}_i)^2}{\sigma_{\text{RV},i}^2 + s^2} \\
    + \ln (2\pi(\sigma_{\text{RV},i}^2 + s^2))\Bigg],
\end{multline}
\noindent where $\text{RV}_{\text{pred}}(t_i)$ is calculated using the standard Keplerian orbit equations \citep{Binnendijk60}. The astrometric orbit term is
\begin{multline}
    \ln \mathcal{L} = -\frac{1}{2} \sum_{i=1}^{N} \left[\frac{1}{s_{\text{ast}}^2} \mathbf{r}_i^\mathrm{T} \, \mathbf{C}_i^{-1} \, \mathbf{r}_i+ \ln \left| \mathbf{C}_i \right|\right] \\
    - 2 N \ln s_{\text{ast}} - N \ln (2 \pi), 
\end{multline}

\noindent where the residuals are
\begin{equation}
    \mathbf{r}_i =
    \begin{bmatrix}
    x_i - x_{\text{pred}}(t_i, \theta) \\
    y_i - y_{\text{pred}}(t_i, \theta)
    \end{bmatrix},
\end{equation}
\noindent and $\mathbf{C}_i$ are the covariance matricies for each measurement. The astrometric orbit model positions $x_{\rm{pred}}(t_i)$ and $\ y_{\rm{pred}}(t_i)$ are
\begin{equation}
    \begin{aligned}
        x_{\rm{pred}}(t) &= x_{\rm{cm}} - A (\cos E - e) - F(1-e^2)^{1/2} \sin E(t) \\
        y_{\rm{pred}}(t) &= y_{\rm{cm}} - B (\cos E - e) - G(1-e^2)^{1/2} \sin E(t), 
    \end{aligned}
\end{equation}
\noindent where $E$ is the eccentric anomaly, and $x_{\rm{cm}}$ and $y_{\rm{cm}}$ are the center-of-mass of the orbit. Since we are modeling the orbit in the frame of the photometric primary, $x_{\rm{cm}} = y_{\rm{cm}} = 0$. The coefficients $A$, $B$, $F$, and $G$ are the Thiele-Innes elements computed from the standard Campbell orbital elements:
\begin{equation}
    \begin{aligned}
        A &= a_0 \left(\cos\Omega \cos\omega - \cos i \, \sin\Omega \sin\omega \right) \\
        B &= a_0 \left(\sin\Omega \cos\omega + \cos i \, \cos\Omega \sin\omega \right) \\
        F &= a_0 \left(-\cos\Omega \sin\omega - \cos i \, \sin\Omega \cos\omega \right) \\
        G &= a_0 \left(-\sin\Omega \sin\omega + \cos i \, \cos\Omega \cos\omega \right).
    \end{aligned}
\end{equation}

\begin{figure*}
    \centering
    \includegraphics[width=0.95\linewidth]{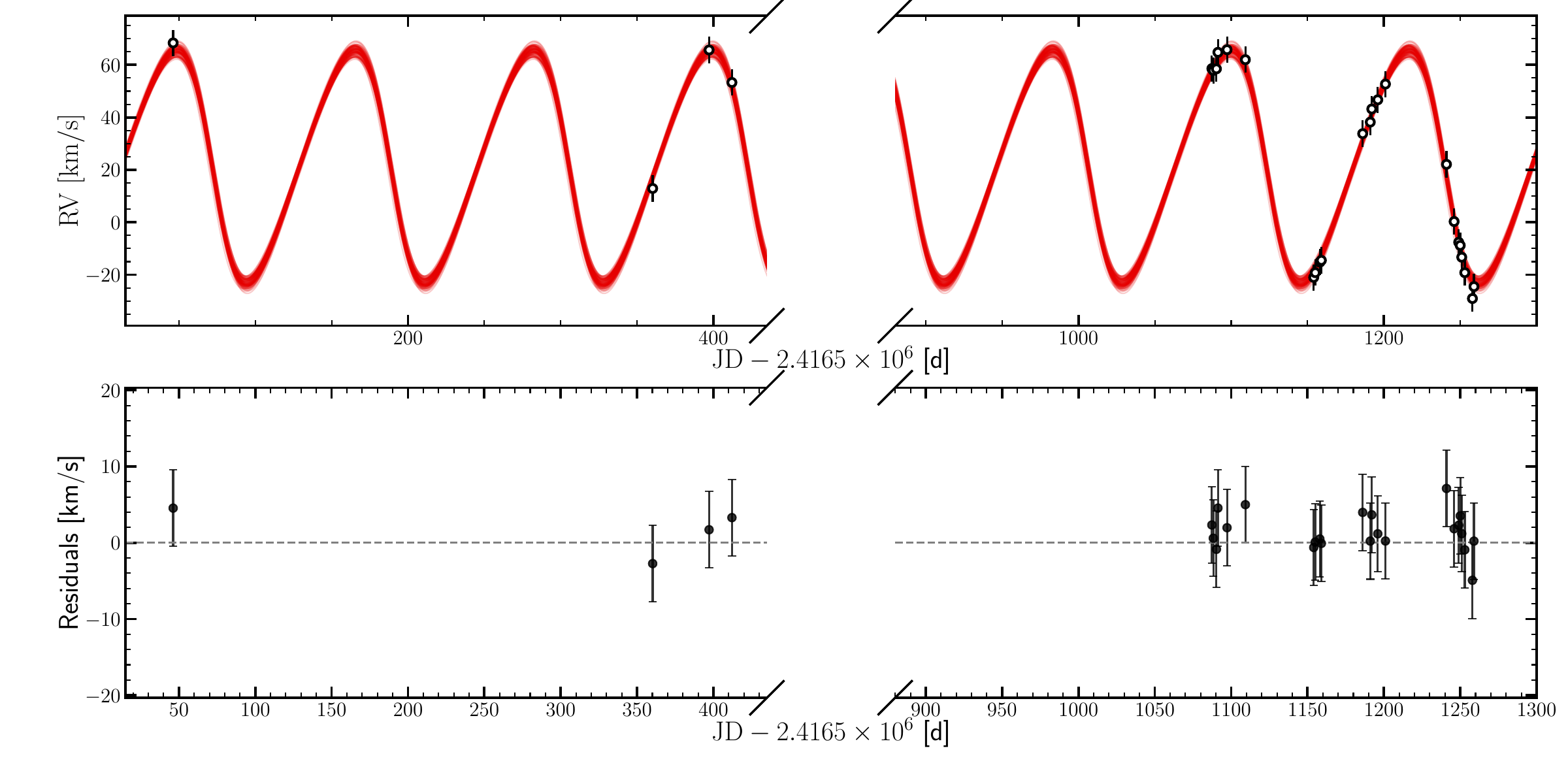}
    \includegraphics[width=0.95\linewidth]{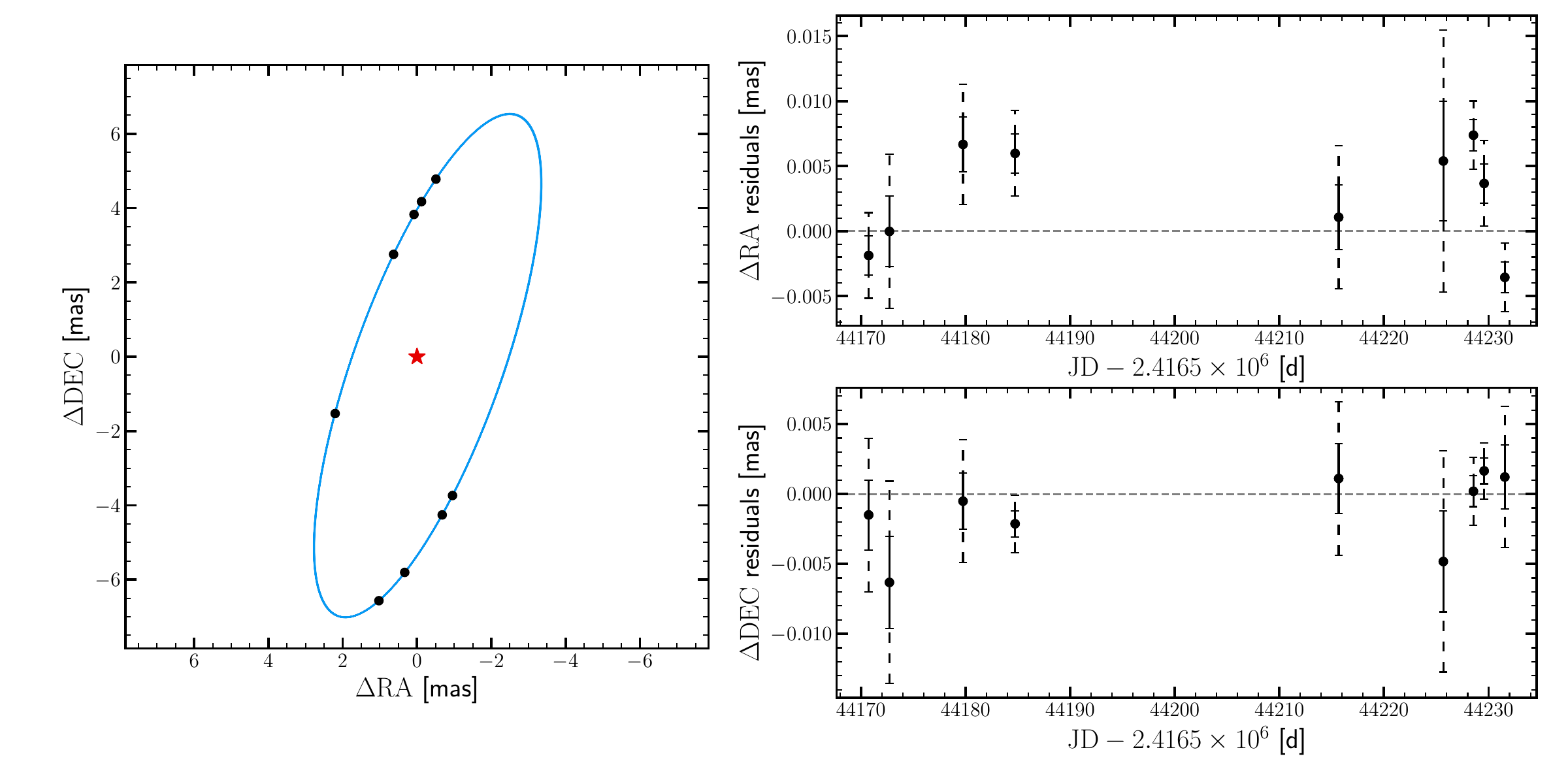}
    \caption{Top: RV orbit model and residuals. The red lines show random samples from the MCMC posteriors. Bottom left: VLTI/GRAVITY astrometry measurements (black points) and random samples for the astrometric orbit model (blue). Bottom right: residuals for the astrometric orbit model. The solid errorbars give the uncertainty derived from the PMOIRED model, and the dashed errorbars report the scaled uncertainty $\sigma_i^\prime = s_{\text{ast}} \sigma_i$.}
    \label{fig:orbit_model}
\end{figure*}

\begin{figure*}
    \centering
    \includegraphics[width=\linewidth]{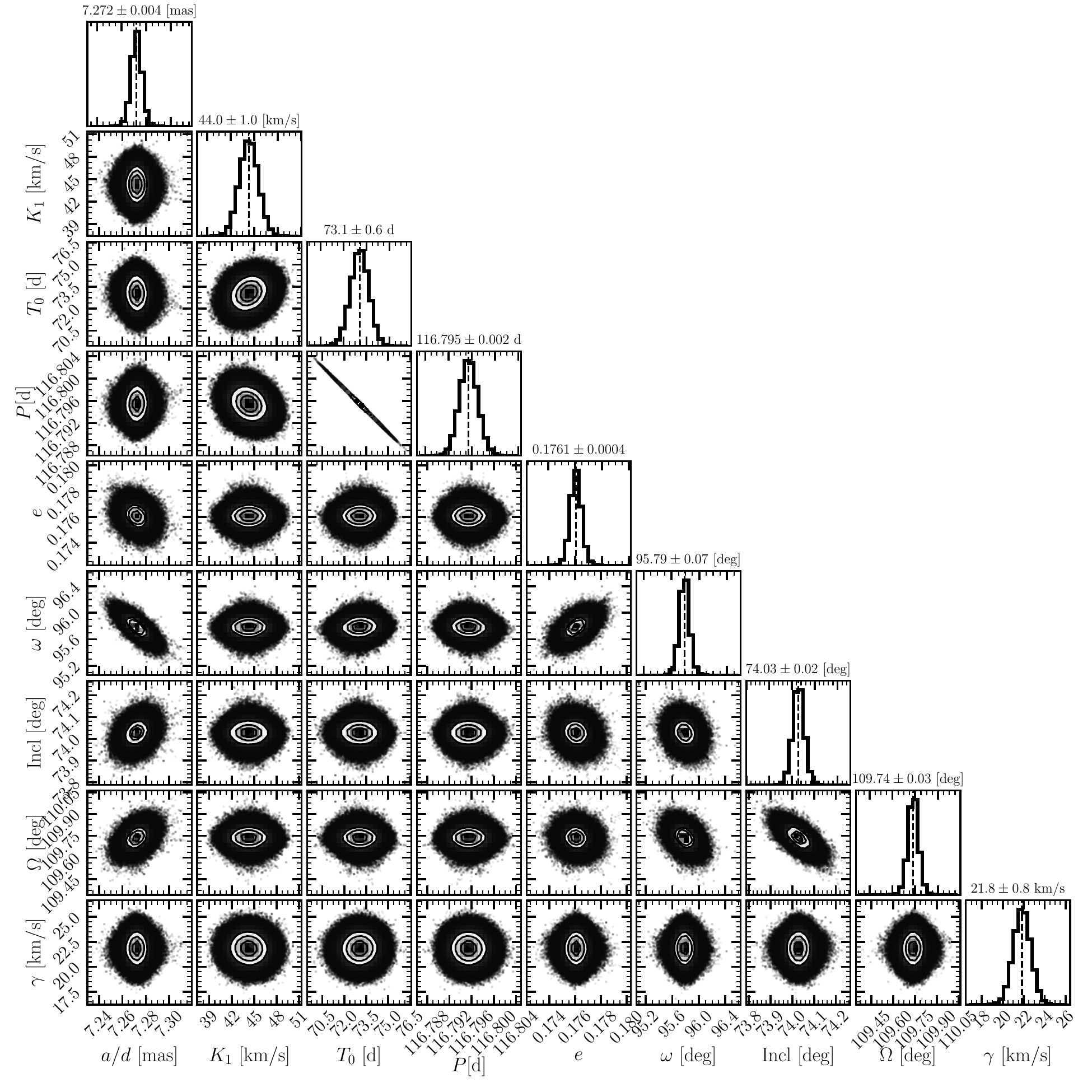}
    \caption{MCMC posteriors for orbital parameters from the simultaneous fit to the RVs and the relative astrometry.}
    \label{fig:rv_ast_corner}
\end{figure*}

We use 20 walkers and run the chains for 500{,}000 iterations. We visually inspect the walker distributions and discard the first 100{,}000 iterations as burn-in. We find that the Gelman-Rubin statistic is $\hat{R} < 1.01$ and the effective sample size is $>10{,}000$ for all parameters, suggesting that the chains have converged.
Figure \ref{fig:orbit_model} shows the resulting orbit model and Figure \ref{fig:rv_ast_corner} shows the corner plot of the MCMC posteriors and random samples of the orbit model compared to the VLTI astrometry observations. Table \ref{tab:orbit_table} reports the median and $1\sigma$ uncertainties of the posteriors. 

\begin{figure}
    \centering
    \includegraphics[width=\linewidth]{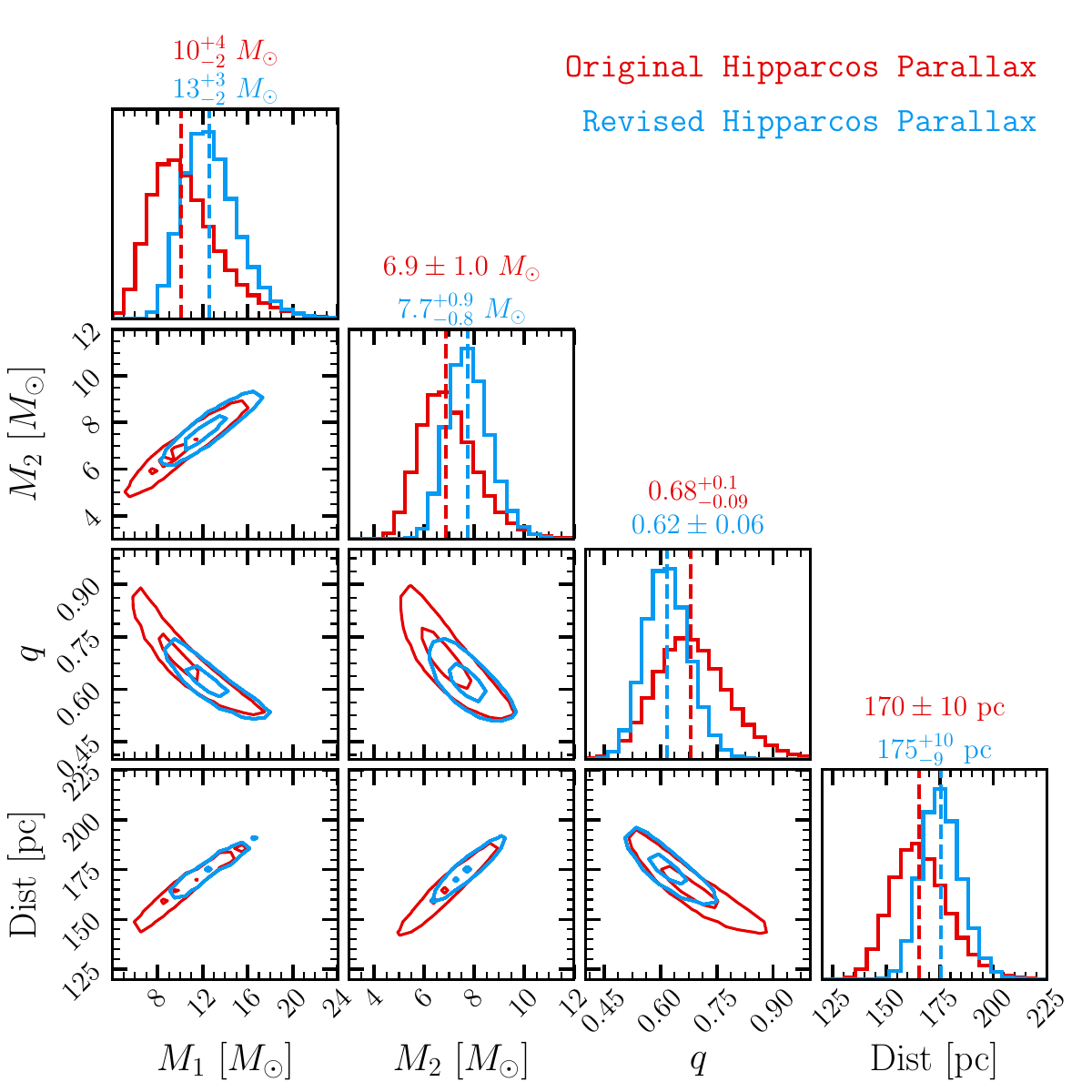}
    \caption{Mass and mass ratio posteriors computed using the orbit posteriors (Figure \ref{fig:rv_ast_corner}) and the original (red) and revised (blue) Hipparcos parallax measurements. Both distance measurements are consistent with the binary being a massive B+B star system. }
    \label{fig:corner_masses}
\end{figure}

Given a parallax, $\varpi$, and associated uncertainty, $\sigma_\varpi$, the binary masses $M_1$ and $M_2$ can be solved for using Kepler's third law. Figure \ref{fig:corner_masses} shows the mass posteriors using the original Hipparcos parallax $\varpi=6.05\pm0.48$~mas \citep{Esa97_publication}. The revised Hipparcos parallax, which primarily includes improvements to attitude reconstruction, is $\varpi=5.7\pm0.3$~mas \citep{vanLeeuwen07}. The revised Hipparcos astrometry is a ``stochastic'' solution, rather than the standard five-parameter solution, which includes an additional ``cosmic noise'' astrometric noise term. Unfortunately, although \kapvel{} has been observed by \Gaia{}, since it is $G=2.5$~mag, it is likely saturated \citep{Fabricius21} and has no reported parallax. 

Using the original Hipparcos parallax, we measure masses of $M_1 = 10^{+4}_{-2}\ M_\odot$ and $M_2 = 6.9\pm1.0 \ M_\odot$ for the photometric primary and secondary, respectively, consistent with a binary system made up of two B stars. If we instead use the revised Hipparcos parallax, the masses are  $M_1 = 13^{+3}_{-2}\ M_\odot$ and $M_2 = 7.7^{+0.9}_{-0.8}\ M_\odot$. The mass uncertainties are dominated by the parallax measurement error. 

The orbital motion of the binary system could also bias the measured parallax, adding additional uncertainty to the measured masses. Since the measured mass ratio, $q$, is not the same as the flux ratio, $\epsilon$, the center-of-light, or photocenter, of the system orbits the barycenter with semi-major axis
\begin{equation}
    \alpha = a \left(\frac{q}{1+q} - \frac{\epsilon}{1+\epsilon}\right).
\end{equation}
The ratio between the projected angular size of the photocenter semi-major axis and the measured parallax quantifies the potential bias. For $\alpha$ in AU, this ratio is simply $\alpha \sin i$.

We must estimate the flux ratio of the system in the Hipparcos $H_p$ filter to determine the parallax bias. For a given mass pair of $M_1$ and $M_2$ and metallicity, we interpolate the MIST evolutionary tracks \citep{Dotter16, Choi16} onto the same age grid, calculate the $K$-band flux as a function of stellar age, and use the observed flux ratio to estimate the system age. We use the $H_p$-band flux ratio at this system age to calculate $\alpha \sin i$. We find $\alpha \sin i \approx 0.22$, which indicates that the parallax can be biased by the binary motion by as much as 22\%. Since the mass ratio depends on the parallax, this fractional bias does increase with increasing distance, but only at the percent level. 

\begin{table}
    \centering
    \caption{MCMC posteriors for the combined RV and astrometry orbit fit. We report the mass measurements for the two different Hipparcos parallax measurements.}
    \label{tab:orbit_table}
    \renewcommand{\arraystretch}{1.5}
    \begin{tabular}{lrr}
\hline
Parameter & \multicolumn{2}{c}{Value} \\
\hline
$a/d\ \rm{[mas]}$ & \multicolumn{2}{c}{$7.272\pm0.004$} \\
$K_1\ \rm{[km/s]}$ & \multicolumn{2}{c}{$44.0\pm1.0$} \\
$P \rm{[d]}$ & \multicolumn{2}{c}{$116.795\pm0.002$} \\
$T_0\ \rm{[d]}$ & \multicolumn{2}{c}{$73.1\pm0.6$} \\
$e$ & \multicolumn{2}{c}{$0.1761\pm0.0004$} \\
$\omega\ [\rm{deg}]$ & \multicolumn{2}{c}{$95.79\pm0.07$} \\
$\rm{Incl}\ [\rm{deg}]$ & \multicolumn{2}{c}{$74.03\pm0.02$} \\
$\Omega\ [\rm{deg}]$ & \multicolumn{2}{c}{$109.74\pm0.03$} \\
$\gamma\ \rm{[km/s]}$ & \multicolumn{2}{c}{$21.8\pm0.8$} \\
$\ln(s\ \rm{[km/s]})$ & \multicolumn{2}{c}{$-5.0\pm3.0$} \\
$\ln(s_{\rm{ast}}\ \rm{[mas]})$ & \multicolumn{2}{c}{$0.8\pm0.2$} \\
$\delta\rm{RV}_{0}\ \rm{[km/s]}$ & \multicolumn{2}{c}{$1.0\pm1.0$} \\
\hline
& $\varpi = 6.05 \pm 0.48$ mas & $\varpi = 5.7 \pm 0.3$ mas \\
\hline
$M_1\ [M_\odot]$ & $10^{+4}_{-2}$ & $13^{+3}_{-2}$ \\
$M_2\ [M_\odot]$ & $6.9\pm1.0$ & $7.7^{+0.9}_{-0.8}$ \\
$q$ & $0.68^{+0.1}_{-0.09}$ & $0.62\pm0.06$ \\
\hline
\end{tabular}

\end{table}

\section{UVES Spectra} \label{sec:uves}

\kapvel{} was previously observed with VLT UVES as part of the ESO Diffuse Interstellar Bands Large Exploration Survey \citep[EDIBLES,][194.C-0833]{Cox17}. \kapvel{} was observed on 2017-04-18 in all four UVES cross-dispersers (CDs). The Phase 3 data product includes the combined spectra for 20 individual exposures in CD1 and CD2 and for 14 and 15 exposures in CD3 and CD4, respectively. While the purpose of the EDIBLES program was to characterize diffuse interstellar bands (DIBs), we can also use the UVES \kapvel{} spectrum to characterize the photometric primary of the binary. We start by continuum normalizing the CD2 spectrum with a cubic spline, which covers a wavelength range of 370--500nm. We mask the Balmer lines during continuum normalization. Our analysis only uses the CD2 observations given the density of spectral features and lack of telluric lines. 

We construct a grid of synthetic B star spectra using TLUSTY model atmospheres from the precomputed BSTAR2006 grid \citep{Lanz07}. We compute synthetic spectra with {\tt Synspec} using these model atmospheres \citep{Hubeny11}. The grid is ranges from $15 \leq T_{\rm{eff}} \leq 30$~kK with a step size of $1$~kK and $2.5 \leq \log g \leq 4.5$ with a step size of $0.25$. We generate the grid at three metallicities: $Z=Z_\odot/2,\ Z_\odot,\ \rm{and}\ 2Z_\odot$. The microturbulence is fixed at 2~km/s since hot stars ($T_{\rm{eff}} \gtrsim 10{,}000$~K) are not expected to have large microturbulent velocities \citep{Landstreet09}. The synthetic spectra are broadened to match the UVES resolution ($R\approx 71{,}000$) with a Gaussian kernel. 

\begin{figure*}
    \centering
    \includegraphics[width=.9\linewidth]{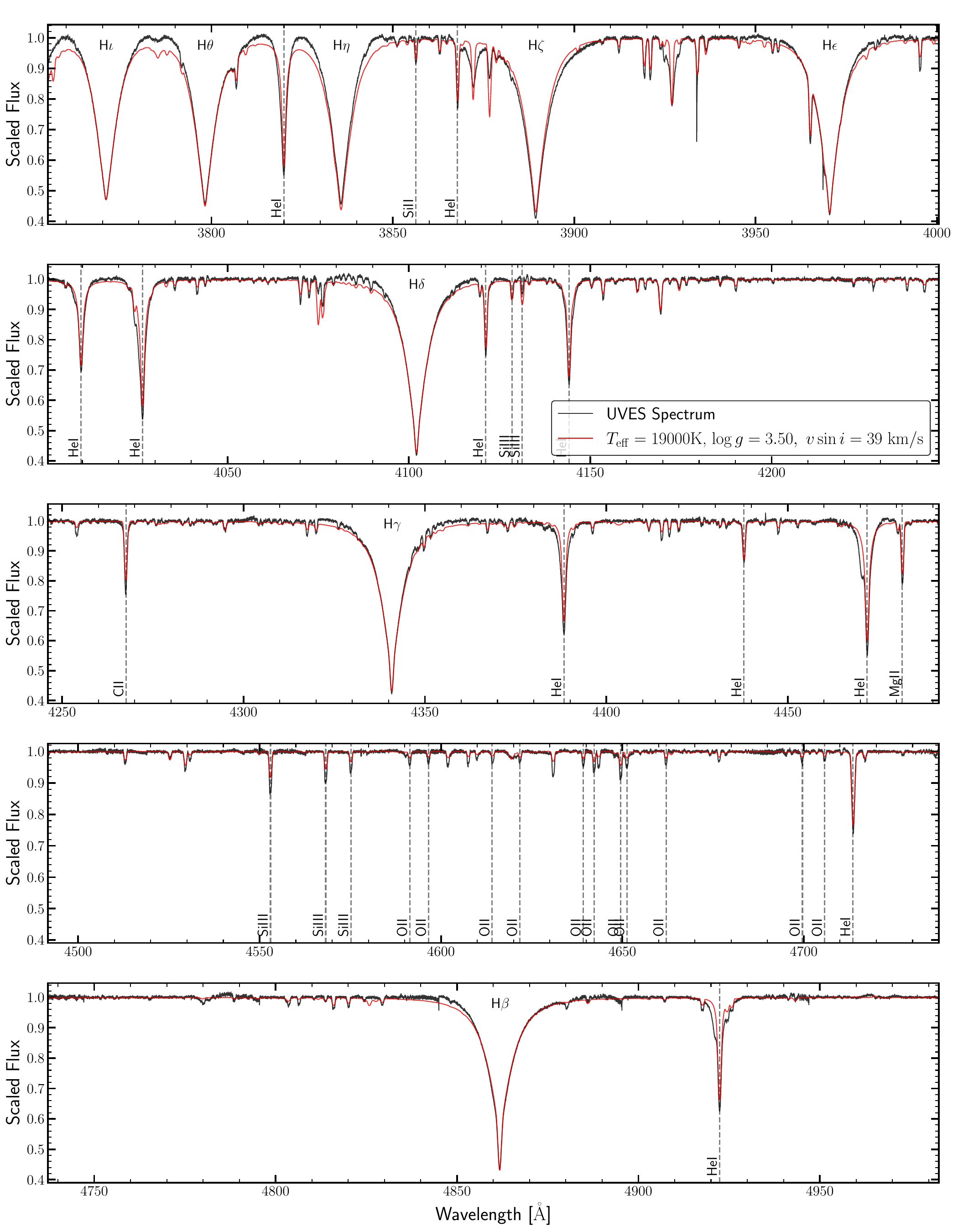}
    \caption{High-resolution UVES spectrum (black) taken as part of the EDIBLES survey \citep{Cox17}. The red line shows the best fitting synthetic spectrum computed using TLUSTY model atmospheres and {\tt Synspec}. Hydrogen, helium, and metal lines are marked using wavelengths from the NIST database \citep{NIST_ASD}.}
    \label{fig:uves}
\end{figure*}

For each synthetic template spectrum, we use $\chi^2$ minimization with Nelder-Mead to determine the RV and the projected rotational velocity, $v \sin i$. We rotationally broaden the spectrum using the method described in \citet{Carvalho23}. We do the $\chi^2$ minimization in two steps. First, we optimize for $v \sin i$ and the RV in $10$~\AA{} windows around the \ion{He}{1} lines (Figure \ref{fig:uves}), which are more sensitive to rotational broadening. Then, we fix the $v \sin i$ at the best-fit value and refine the RV with $\chi^2$ minimization using the full UVES CD2 wavelength range. We select the template with the lowest minimized $\chi^2$ as our best-fitting template. 


We find the best match at $T_{\rm{eff}} = 19$~kK, $\log g=3.50$ and $Z = 0.5 Z_\odot$. At Solar metallicity, we find the same best-matching effective temperature and surface gravity. The measured effective temperature and surface gravity are consistent with estimates from broadband Str\"{o}mgren $ubvy$ photometry \citep{Leone97}. For the best fitting template, we measure $\rm{RV}=26.5$~km/s and $v \sin i = 39$~km/s. This is the radial velocity we use when fitting the combined orbit model (Section \S\ref{sec:orbit}). The projected rotational velocity is lower than those of B-stars found in open clusters \citep[e.g.,][]{Santos25}, but consistent with B-stars found in the Galactic field \citep{Huang08}. 

Figure \ref{fig:uves} shows the UVES spectrum and the synthetic spectrum of the best-fitting B-star model and marks hydrogen Balmer, \ion{He}{1} and metal lines. The synthetic spectrum generally reproduces the observed \ion{He}{1} and metal lines, but less accurately matches the wings of the hydrogen Balmer lines, especially at $\lambda < 4000$\,\AA{}. This is likely to represent an artifact of the continuum normalization procedure in areas of the spectrum dominated by the broad Balmer line wings. The \ion{Si}{2} and \ion{O}{2} lines between $4550 \lesssim \lambda \lesssim 4700$~\AA{} are also under predicted by the best matching synthetic template, which could suggest that \kapvel{} is alpha-enhanced. This analysis also assumes that the spectrum is entirely dominated by the photometric primary. We evaluate this assumption in more detail in Section \S\ref{sec:evotracks}.

\begin{figure}
    \centering
    \includegraphics[width=\linewidth]{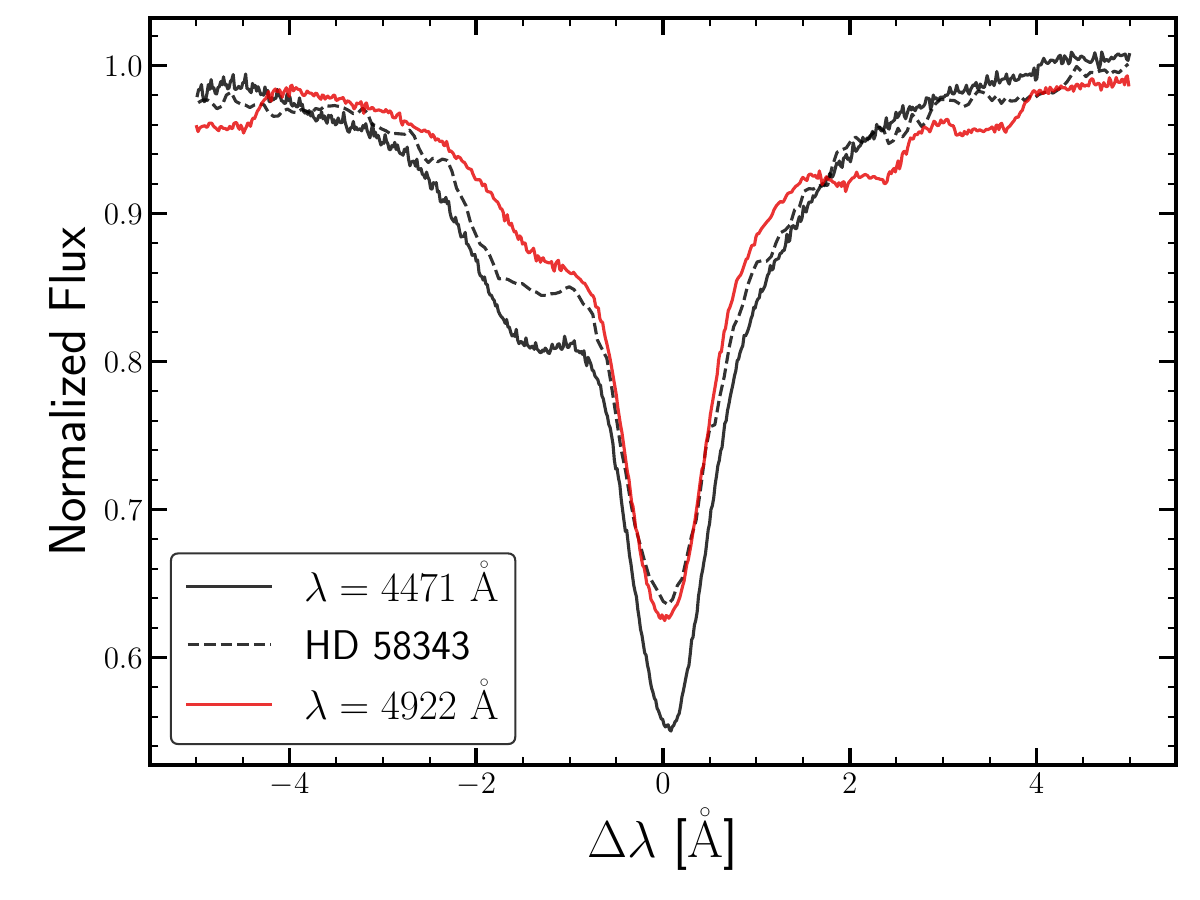}
    \caption{UVES spectrum of \kapvel{} showing the two \ion{He}{1} absorption lines with asymmetric blue wings not predicted by the synthetic spectra. The $\lambda$4471 line of the Be2V star HD 58343 from \citet{Chauville01} is shown for comparison. We find that the apparent second component in the blue wings of these absorption lines is more likely from forbidden [\ion{He}{1}] transitions rather than the stellar secondary.}
    \label{fig:he_lines_zoom}
\end{figure}

The spectrum is consistent with a B2V-IV star. In addition to having no \ion{He}{2} lines, the \ion{Si}{2} doublet at $\lambda$4129 is clearly visible, and the \ion{C}{2} $\lambda$4267 line is strong. The spectrum is qualitatively similar to the B2V spectroscopic standard $\beta^2$~Sco \citep{Negueruela24}. The most notable differences between the best matching synthetic spectrum and the UVES spectrum of \kapvel{} are the \ion{He}{1} lines at $\lambda\lambda$4471, 4922. Figure \ref{fig:he_lines_zoom} shows a zoom in on these absorption lines. In both cases, we observe an additional blended component in the blue wing of the primary absorption feature that is not represented by our synthetic spectra at any effective temperature, surface gravity, or metallicity. We consider two possibilities: (a) the additional component is actually the \ion{He}{1} absorption line of the fainter B-star companion or (b) the line list we use with {\tt Synspec} does not include this blended absorption line. 

For the former scenario, we find that if the component in the blue wing is the \ion{He}{1} of the companion, the velocity difference between the components would be $\sim 90$~km/s. Since the UVES spectrum was taken near conjunction, we expect the velocity difference between the two components to be small. An unrealistically small mass ratio $q \lesssim 0.07$ would be needed to produce a $\Delta\rm{RV} = 90$~km/s between the binary components at the epoch of the UVES. Furthermore, we do not see evidence for blended components in the other \ion{He}{1} ines (e.g., $\lambda$4120, $\lambda$4143, $\lambda$4387) or in the \ion{C}{2} or \ion{Mg}{2} lines, so we reject this scenario.

While the structure in the blue wings of the \ion{He}{1} lines at 4471\AA{} and 4922\AA{} are not predicted by our synthetic spectra, similar \ion{He}{1} lines are found in other B and Be stars. For example, \citet{Negueruela24} show that the B3V star HD 178849 has structure in the \ion{He}{1} $\lambda$4471 line that is not apparent in stars of the same spectral class with higher rotational velocities. Similarly, the spectrum of HD~58343, a Be2V star with $v \sin i = 43$~km/s also has a similar $\lambda$4471 line profile \citep{Chauville01} which we show in Figure \ref{fig:he_lines_zoom} for comparison. The most likely explanation is that the blue component of the $\lambda$4471 comes from the forbidden transition [\ion{He}{1}] at $\lambda$4469.96 \citep{Daflon07}. This is is less commonly observed because it requires a high-resolution spectrum and the star to have projected rotational velocity $v \sin i \lesssim 50$~km/s (see Figure 6 of \citealt{Santos25}). A forbidden [\ion{He}{1}] line is also expected at $\lambda$4921 \citep{Beauchamp98}, accounting for the blue wing in the \ion{He}{1} $\lambda$4922 profile seen in Figure \ref{fig:he_lines_zoom}. 

\section{Comparison to Evolutionary Tracks} \label{sec:evotracks}

\begin{figure}
    \centering
    \includegraphics[width=\linewidth]{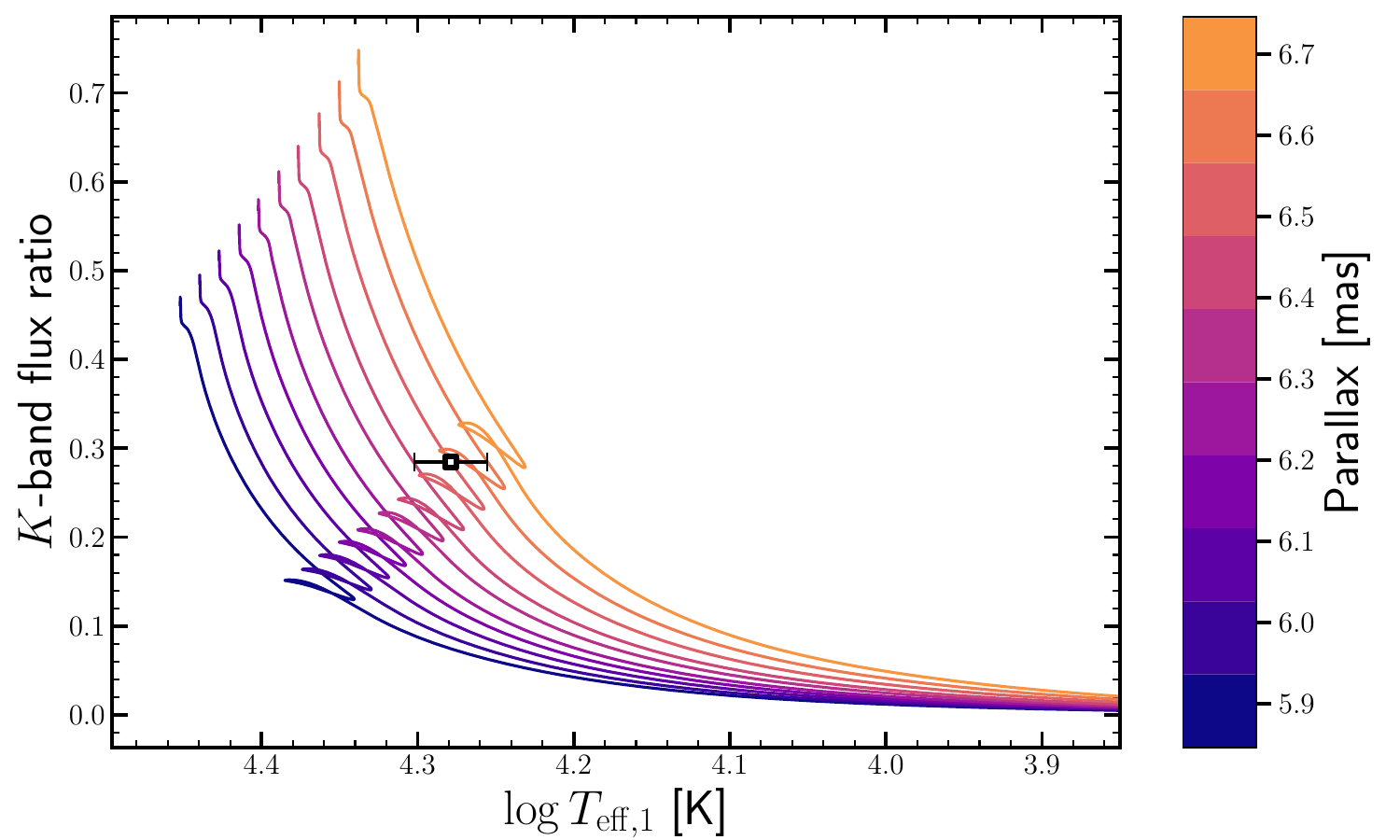}
    \caption{MIST evolutionary tracks for binary masses derived from the astrometric+RV orbit model. Since the measured masses depend on the distance to the binary, the different colors show the tracks for different parallaxes. The flux ratio from the VLTI/GRAVITY observation and the effective temperature from the UVES spectrum (black marker) are consistent with a parallax of $6.5$~mas.}
    \label{fig:evotracks}
\end{figure}

By combining the measured binary masses and the flux ratio from the VLTI/GRAVITY observations, we can estimate the evolutionary state of the binary and the spectroscopic properties of the secondary. We download precomputed evolutionary tracks \citep{Dotter16, Choi16} from MESA Isochrones and Stellar Tracks. For a given mass pair of $M_1$ and $M_2$ and metallicity, we interpolate their tracks onto the same age grid and calculate the $K$-band flux as a function of stellar age. We can then estimate the properties of both stars and the age of the system.

The measured masses are strongly dependent on the assumed parallax (Figure \ref{fig:corner_masses}). The parallax uncertainty is large ($0.3$ and $0.48$~mas for the original and updated measurement, respectively) and the parallax could be biased by the orbital motion by as much as 22\% (Section \S\ref{sec:orbit}). The effective temperature of the primary star measured from UVES can be used to place additional constraints on the masses of the stars. Figure \ref{fig:evotracks} shows the evolution of the flux ratio of the binary system as a function of the effective temperature of the primary star assuming different parallaxes. We assume a metallicity $[\rm{Fe/H}] = -0.3$ based on our template minimization results. The measured flux ratio, $\epsilon = 0.285 \pm 0.003$, implies dramatically different primary effective temperatures depending on the parallax and the derived primary mass. For example, assuming the original Hipparcos parallax $\varpi = 6.05$~mas, the flux ratio measurement implies that the photometric primary is still about halfway through its main sequence lifetime and has an effective temperature of $T_{\rm{eff}, 1} > 25$~kK, which is much hotter than the temperature we estimate from the UVES spectrum. Figure \ref{fig:evotracks} shows that the effective temperature estimate from UVES and binary flux ratio prefer a parallax of $\varpi \approx 6.5$~mas. This is $\gtrsim 1\sigma$ larger than the original Hipparcos parallax measurement and $\gtrsim 3\sigma$ larger than the revised Hipparcos parallax measurement using the reported uncertainties, but within the broader range of parallaxes acceptable if the measurement has been biased by the binary motion. 

Assuming this value for the parallax, we can assess whether or not we expect to see any spectral features of the secondary in the UVES spectrum. If $\varpi \approx 6.5$~mas, the primary and secondary masses are $M_1 = 7.5\ M_\odot$ and $M_2 = 5.9\ M_\odot$, respectively. We use these masses and the orbital solution and predict that velocity difference between the primary and the secondary star at the time of the UVES observation is only $\Delta \rm{RV} = 12.2$~km/s, which is not surprising given that the UVES observation was taken near conjunction. At a system age of $\sim 40$~Myr, which is when the $K$-band flux ratio matches the VLTI/GRAVITY result, both stars have $T_{\rm{eff}}\approx 19{,}000$~K, so the flux ratio is roughly the same in the $K$-band and in the wavelength range of the UVES spectrum. We take the synthetic spectra from the TLUSTY BSTAR2006 grid described in Section \S\ref{sec:uves} and select the closest matches to the primary and secondary star. We broaden these to match the UVES resolution and apply the expected RV shifts for each component. We also rotationally broaden the spectra of both components by 40~km/s. We combine the synthetic primary and secondary spectra, weighting them such that the flux ratio is $\epsilon = 0.28$. We find that the differences between the spectrum of the photometric primary and the combined binary spectrum are at the $\lesssim 2\%$ level. It is therefore unsurprising that we do not detect any signatures of the secondary in the UVES spectrum. 

\begin{figure*}
    \centering
    \includegraphics[width=0.9\linewidth]{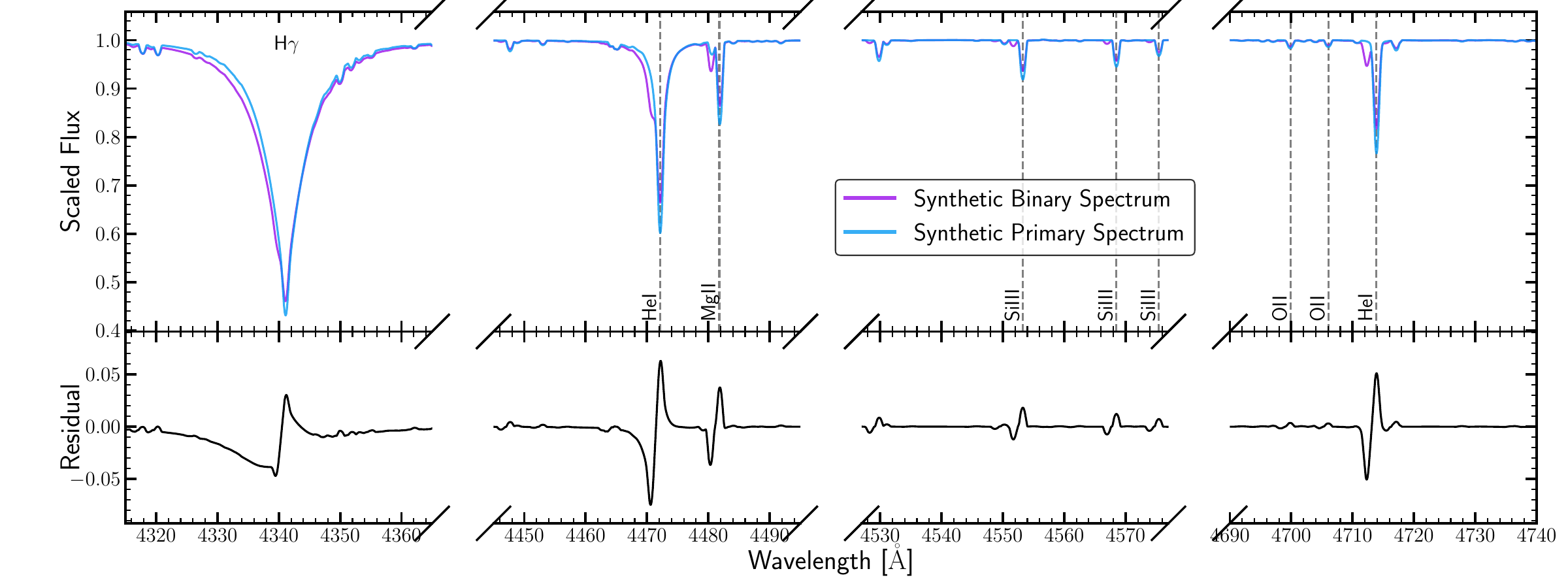}
    \caption{Synthetic spectrum of the binary system at RV quadrature, when the velocity difference between the two stars is the largest and the secondary star is mostly likely to be detected spectroscopically. The spectrum of the flux-weighted combination of the binary components (purple) differs most dramatically from that of the primary's spectrum (blue) in the wings of the Balmer absorption lines and will also produce double-peaked \ion{He}{1} and some metal lines, like \ion{Si}{2}. The bottom panel shows that, at RV quadrature, the two spectra differ at the $\sim 5\%$ level. We note that the actual archival UVES spectrum (Figure \ref{fig:uves}) was taken near conjunction.}
    \label{fig:quad_spec}
\end{figure*}

We also simulate the binary spectrum at RV quadrature, where we would be most likely to detect the spectroscopic signatures of the fainter companion. Figure \ref{fig:quad_spec} compares the synthetic spectrum of the primary at quadrature to the flux-weighted combined spectrum of both components. We find that, at quadrature, the spectrum of the primary and the combined binary spectrum differ at the $\sim 5\%$ level. The differences are most apparent in the asymmetries in the wings of the Balmer lines and split He~I and metal lines. 

\section{Discussion and Conclusions} \label{sec:discussion}

\kapvel{} is one of the brightest stars in the Southern sky and was first detected as a single lined spectroscopic binary by \citet{Curtis1907}. Here, we use VLTI/GRAVITY to definitively determine the nature of the secondary star for the first time. To the best of our knowledge, \citet{Daszynska17} present the only previous evidence that the companion is a massive star and not a compact object. They used BRITE photometry to characterize $g$-mode oscillations and found a frequencies inconsistent with the spectral type and evolutionary state of the B2V-IV photometric primary. They instead suggest that the oscillations come from a fainter, less massive B star companion. 

We observed \kapvel{} star as part of a broader effort to identify stars with compact object companions. Given the predicted large angular separation (Figure \ref{fig:candidate_selection}) and large companion mass (Figure \ref{fig:companion_mass}), \kapvel{} emerged as a promising candidate for VLTI/GRAVITY observations. Although were were unambiguously able to identify the luminous secondary in the first VLTI observation, we obtained a total of nine epochs, and simultaneously fit the astrometry and RVs to model the binary orbit (Figure \ref{fig:orbit_model}). From this model, we can predict the masses of the binary components (Figure \ref{fig:corner_masses} and Table \ref{tab:orbit_table}). We find that the uncertainty in the mass measurement is primarily driven by the uncertainty in the Hipparcos parallax. It is also likely that the binary motion has biased the parallax measurement by as much as $\sim 22\%$. 

Given its spectral type and magnitude, \kapvel{} has also been targeted by studies of diffuse interstellar bands. We use the publicly available UVES spectrum from \citet{Cox17} and B-star model atmospheres to estimate the effective temperature, surface gravity, and rotational velocity of the photometric primary (Figure \ref{fig:uves}). By combining the binary flux ratio measured from VLTI/GRAVITY and the effective temperature of the photometric primary, we can estimate the distance and binary masses (Figure \ref{fig:evotracks}). 

We also use this model to estimate how the secondary star contributes to the combined spectrum. Even at RV quadrature, when the velocity difference between the two stars is largest, the synthetic binary system is expected to differ from the photometric primary at the few-percent level (Figure \ref{fig:quad_spec}). While it may be possible to detect the subtle spectroscopic signatures of the secondary with multi-epoch high-resolution spectra and techniques like spectral disentangling \citep[e.g.,][]{Shenar20, Seeburger24}, we detect the luminous secondary with the first VLTI/GRAVITY observation. Since the orbital period and ephemeris is known from the RVs, a single VLTI observation could be used to rule out a black hole with a total observing time of $\sim 1$~hour, including overheads. For comparison, the single UVES spectrum shown in Figure \ref{fig:uves} had a total observing time of 23~min, and multiple spectra ($\gtrsim 5$) taken at different orbital phases are typically needed for spectral disentangling to be effective. 

\subsection{Other VLTI Targets}

\begin{table*}
    \centering
    \caption{Other candidates for VLTI/GRAVITY imaging selected from the SB9 catalog (top part of table) and \Gaia{} DR3 (bottom). Archival observations confirm that most of the SB9 targets are regular stellar binaries. Most of the \Gaia{} DR3 candidates are in the Hertzsprung gap and are likely luminous binaries with one red and one blue component.}
    \sisetup{table-auto-round,
     group-digits=false}
    \setlength{\tabcolsep}{2pt}
    \renewcommand{\arraystretch}{1.0}
    \begin{tabular}{l r r r r r}
\toprule
{Name} & {$K_s$} & {Period} & {Projected Sep.} & {$M_{2,\rm{min}}$} & {Notes} \\
{} & {[mag]} & {[d]} & {[mas]} & {$[M_\odot]$} & {}\\ 
\midrule
X Per & 5.9 & 3900.0 & 6.6 & 20.7 & Be X-ray binary pulsar \citep{Fortin23} \\
HD 45910 & 4.4 & 232.5 & 1.7 & 8.1 & B2+KIII \citep{Cowley64} \\
$\chi$ Lup & 4.0 & 15.3 & 4.1 & 7.2 & B9.5V+ A2V \citep{LeBouquin13} \\
$\delta$ Gem & 2.6 & 2238.6 & 353.1 & 6.0 & Uncertain RV orbit \citep{Abt74} \\
HD 187299 & 3.5 & 1901.0 & 7.4 & 5.9 & G5I + B7V \citep{Griffin79} \\
HD 50975 & 5.7 & 190.2 & 0.8 & 5.2 & F8I + B2 \citep{Sperauskas14} \\
HD 36881 & 5.0 & 1857.0 & 12.7 & 4.3 & VLTI P116 Candidate \\
HD 8556 & 4.9 & 5895.0 & 240.3 & 4.1 & F0 + F5 \citep{Fletcher73} \\
71 Tau & 4.0 & 5200.0 & 227.4 & 3.9 & HST UV emission lines \citep{Simon00} \\
HD 127208 & 5.1 & 24.6 & 1.0 & 3.8 & F + Be \citep{Dempsey90} \\
$\omega$ Eri & 3.7 & 3057.0 & 112.5 & 3.7 & RV constant \citep{Merle24} \\
HD 70826 & 5.5 & 3900.0 & 25.7 & 3.5 & G7III + (A+A) \citep{Carquillat05} \\
HIP 39794 ($\epsilon$ Vol) & 4.7 & 14.2 & 1.2 & 3.2 & B6IV + B8 \citep{Veramendi14} \\
HIP 30891 (V723 Mon) & 5.4 & 59.9 & 1.1 & 3.2 & Stripped giant + subgiant \citep{ElBadry22_zoo} \\
HIP 51213 (45 Leo) & 6.0 & 12658.0 & 146.2 & 3.2 & Uncertain RV orbit \\
HIP 23900 & 5.3 & 58.3 & 1.4 & 3.1 & B2V + B5-B8V \citep{Tarasov16} \\
HIP 99853 (22 Vul) & 2.9 & 249.1 & 3.3 & 3.0 & Eclipsing G3I + B8V \citep{Griffin93} \\
\hline
Gaia DR3 5824739062379517824 & 8.4 & 1113.3 & 7.7 & 40.1 & B dwarf with uncertain long-period \Gaia{} orbit \\
Gaia DR3 5857059996952633984 & 9.1 & 155.1 & 0.6 & 8.2 & Hertzsprung Gap; \TESS{} eclipsing binary\\
Gaia DR3 3331748140308820352 & 9.1 & 225.3 & 0.5 & 5.7 & Hertzsprung Gap \\
Gaia DR3 5971649891874353536 & 8.5 & 50.0 & 0.6 & 5.1 & Hertzsprung Gap; \TESS{} eclipsing binary\\
Gaia DR3 2021374066702077312 & 7.6 & 44.4 & 0.5 & 4.0 & Hertzsprung Gap \\
Gaia DR3 5307540956005282560 & 7.7 & 212.7 & 0.9 & 3.6 & Hertzsprung Gap; \TESS{} eclipsing binary\\
Gaia DR3 5879279305887191552 & 7.7 & 167.7 & 0.6 & 3.4 & Hertzsprung Gap \\
Gaia DR3 6021285355771958528 & 9.0 & 269.5 & 6.3 & 3.3 & -- \\
Gaia DR3 3179591300281505664 & 9.2 & 857.7 & 2.0 & 3.2 & Hertzsprung Gap \\
Gaia DR3 2017710455239591424 & 9.0 & 331.7 & 1.1 & 3.2 & -- \\
Gaia DR3 6077083852882264320 & 8.9 & 897.6 & 2.6 & 3.1 & Hertzsprung Gap \\
Gaia DR3 5335226727540945536 & 8.0 & 195.8 & 0.5 & 3.0 & Hertzsprung Gap; \TESS{} eclipsing binary\\
\bottomrule
\end{tabular}

    \label{tab:discussion}
\end{table*}

While \kapvel{} does not host a black hole companion, the VLTI/GRAVITY observations of this system have demonstrated how the combination of RVs and high-contrast imaging may be used to effectively and efficiently vet other candidate star+black hole binaries. As described in Section \S\ref{sec:targetselection}, we selected candidates starting from the SB9 catalog \citep{Pourbaix04} and the \Gaia{} single-lined spectroscopic binary catalog. There are $\sim 40$ other binaries with minimum companion mass $M_{2, \rm{min}} > 3\ M_\odot$ and projected angular separations $>0.5$~mas that are observable with VLTI.

Table \ref{tab:discussion} reports the properties of these targets. We reject the overwhelming majority of star+black hole candidates based on archival observations, including previous spectroscopic follow-up, speckle imaging, and direct imaging. In some cases, we also use photometry from \textit{TESS} and identify eclipses. There are four candidates from the SB9 catalog where the luminous companion has not been previously observed: 

\textbf{$\delta$ Gem: }
$\delta$ Gem is a hierarchical triple system \citep{Tokovinin17}. \citet{Abt65} report an orbital period 2238 days for the inner binary, $\delta$~Gem~A. The binary mass function is $3.8\ M_\odot$ and the photometric primary is a F2 main sequence star, so the companion would have to be $\gtrsim 6\ M_\odot$. This orbital solution has since been questioned \citep{Abt74}, and the secondary in the inner binary has never been detected, including through an lunar occultation observation \citep{Tremaine74}. $\delta$ Gem A was also observed with the Navy Precision Optical Interferometer, but \citet{Hutter16} do not report the detection of a companion. This may be a good candidate for high-contrast imaging or interferometry. If a luminous secondary is identified, the star can be immediately rejected as a black hole candidate. If no secondary is found, additional RVs would still be needed to confirm binary orbit.

\textbf{HD 36881: }
HD 36881 is a chemically peculiar B9III star with an orbital period of 1862~d and a binary mass function of $0.85\ M_\odot$. Previous high-resolution spectroscopic and VLT/NACO imaging observations have not identified the companion \citep{Scholler10}. We are observing this system in the current observing period 116 with VLTI/GRAVITY.

\textbf{71 Tau: }
71 Tau is $\delta$ Scuti F0IV-V star with a long orbital period (5200~d) and a large binary mass function, $f(M) = 1.7\ M_\odot$. The companion has not been previously identified with speckle interferometry \citep{Patience98, Mason96}. Like $\delta$~Gem, the orbital ephemeris is uncertain \citep{Abt74}, and additional RVs would be needed to confirm the orbit if no companion is found with high-contrast imaging or interferometry. This star is the second brightest X-ray source in the Hyades cluster, and ultraviolet observations with the International Ultraviolet Explorer found no UV excess indicative of a hot companion \citep{Simon97}. \citet{Simon00} used Hubble Space Telescope UV spectra and identified an emission line source separated by $0\farcs 12$. \citet{Simon00} suggest that this signal is associated with an inner late-type binary companion that is chromospherically active that was undetected in previous interferometry observations that were taken near conjunction. This separation is consistent with the maximum angular separation we predict using \Gaia{} and the SB9 orbital solution (Table \ref{tab:discussion}), but \citet{Bohm07} instead suggest that the offset UV emission lines come from material excited by a supernovae in the cluster. 

\textbf{45 Leo: }
45 Leo is a chemically peculiar B9IV star \citep{Lorden83} with an uncertain long-period orbit with $f(M) = 1.1\ M_\odot$ \citep{Abt73}. The binary has not been resolved with speckle interferometry \citep{Hartkopf84}. 

Table \ref{tab:discussion} also includes the 12 sources selected from \Gaia{} DR3 that are brighter than $K < 9.5$~mag, which is the magnitude limit for single-field mode observations with VLTI/GRAVITY. Almost all of the targets are in the Hertzsprung gap on the color-magnitude diagram, and are likely to be binaries with an evolved red giant star and a massive main sequence star. The evolved star likely dominates the spectral energy distribution in the wavelength range of the \Gaia{} RVS (846--870~nm) so the systems are detected SB1s.  

\Gaia{} DR3 does not include the epoch RVs for the SB1 solutions, only the orbit model and covariance matrix of the posteriors, so it can be challenging to discriminate between accurate binary orbits and spurious solutions \citep[e.g.,][]{Bashi22}. For example, the \Gaia{} SB1 solution for Gaia DR3 5824739062379517824 has a very high mass function, $f(M) = 32^{+6}_{-9}\ M_\odot$, but the orbital period is long ($>1000$~d) and is comparable to the baseline of the Gaia DR3 data. The photometric primary is also an early-type main sequence star with extinction-corrected $G_{\rm{BP}} - G_{\rm{RP}} < 0$. Such a star will likely only have broad hydrogen lines in the \Gaia{} RVS spectrum, so it is possible that the solution is biased by a handful of poor RV measurements. Even though the orbit is uncertain, this may still be a promising target for VLTI/GRAVITY observations since a luminous companion could be identified with a small number of observations taken at different orbital phases. 

While there are some candidates that are suitable for interferometric or high-contrast imaging observations, this search strategy is primarily limited by the limiting magnitude of VLTI/GRAVITY at small angular separations and the number of bright stars with well-characterized long period RV orbits. For a single-field off-axis observation, the limiting magnitude is $K = 9.5$~mag using the ATs. New extreme adaptive optics instruments, such as GRAVITY+ \citep{Gravity25}, will allow us to push towards fainter targets ($K \lesssim 14.5$~mag). Going forward, this strategy might best be used in conjunction with RV follow-up or \Gaia{} astrometry. For example, the catalog of spectroscopic binaries in \citet{Chini12} includes more than 80 SB1s with $K < 8.5$~mag, most of which have less than 10 RVs. By combining spectroscopic monitoring of massive stars \citep[e.g.,][]{Mahy22} with high-contrast imaging or interferometry with instruments such as GRAVITY, SHARK-VIS \citep{Pedichini24}, or CHARA \citep{tenBrummelaar05}, candidates could be vetted more efficiently than with spectroscopic follow-up alone. 

The number of star+compact object candidates is expected to expand dramatically with \Gaia{} Data Release 4, which is scheduled to be released in late 2026 and will include epoch astrometry and radial velocities. \Gaia{} BH3 \citep{GaiaBH3}, which was discovered using pre-release \Gaia{} DR4 astrometry, has already been observed with VLTI/GRAVITY to place limits on the near-infrared emission \citep{Kervella25}. \citet{Nagarajan25} predict that $\sim 30$ black holes will be detectable with \Gaia{} DR4, and some of these long-period candidates will likely be suitable for high-contrast imaging or interferometric observations to confirm dark companions. 

\begin{acknowledgments}

We thank the anonymous reviewer for their useful comments that improved the quality of this manuscript. DMR and TAT thank Chris Kochanek and Kris Stanek for helpful conversations. TAT thanks Keith Smith for motivating questions about the Kappa Velorum system. This research was supported in part by grant NSF PHY-2309135 to the Kavli Institute for Theoretical Physics (KITP).

Support for this work was provided by NASA through the NASA Hubble Fellowship grant HST-HF2-51588.001-A awarded by the Space Telescope Science Institute, which is operated by the Association of Universities for Research in Astronomy, Inc., for NASA, under contract
NAS5-26555.

S.K.\ acknowledges support from an European Research Council (ERC) Consolidator Grant (Grant Agreement ID 101003096), as well as an STFC Small Award (ST/Y002695/1).

The VLTI/GRAVITY observations were collected at the European Organisation for Astronomical Research in the Southern Hemisphere under ESO programme 114.274C.
The UVES analysis is based on observations collected at the European Organisation for Astronomical Research in the Southern Hemisphere under ESO programme 194.C-0833. This research has made use of the SIMBAD database, operated at CDS, Strasbourg, France. This research has made use of the VizieR catalogue access tool, CDS, Strasbourg, France. 

\end{acknowledgments}

\vspace{5mm}




\bibliography{kappavel}{}
\bibliographystyle{aasjournal}

\end{document}